\documentclass[10pt]{article}
\usepackage[a4paper,left=41pt,right=41pt,bottom=60pt,top=60pt]{geometry}

\usepackage{microtype}
\usepackage{fourier}

\twocolumn

\usepackage{color}
\usepackage{cite}
\usepackage{amsmath,amssymb,amsfonts}
\usepackage{algorithmic}
\usepackage{graphicx}
\usepackage{textcomp}
\usepackage{siunitx}[=v2]
\usepackage[version=4]{mhchem}

\newcommand{\zt}{\ensuremath{z_0}}
\newcommand{\intl}{\ensuremath{\int_{0}^{\infty}}}
\newcommand{\smax}{\ensuremath{S_\mathrm{max}}}
\newcommand{\smin}{\ensuremath{S_\mathrm{min}}}
\newcommand{\sfr}{\ensuremath{_\mathrm{FR}}}

\newcommand{\sal}{\ensuremath{_\mathrm{Al}}}
\newcommand{\mean}[2]{\ensuremath{{\left< #1 \right>}_{ #2 }}}

\renewcommand\footnotemark{}

\begin{document}
\title{X-ray dark-field signal reduction due to\\hardening of the visibility spectrum}
\author{
Fabio~De~Marco,
Jana~Andrejewski,
Theresa~Urban,
Konstantin~Willer,
Lukas~Gromann,\\
Thomas~Koehler,
Hanns-Ingo~Maack,
Julia~Herzen,
and Franz~Pfeiffer
\thanks{\hspace*{-1.8em}This work was supported by the European Research Council (ERC) as part of the Horizons 
2020 program under grant AdG~695045, and by the DFG Gottfried Wilhelm Leibniz program.}
\thanks{\hspace*{-1.8em}F.~De~Marco, J.~Andrejewski, K.~Willer, and L.~Gromann were with the Chair of 
Biomedical Physics, Department of Physics, Munich Institute of Biomedical Engineering, Technical University of Munich, 85748~Garching, Germany.
F.~De~Marco is now with the University of Trieste, 32127 Trieste, Italy (e-mail: fabio.de.marco@ph.tum.de).}
\thanks{\hspace*{-1.8em}T.~Urban is with the Chair of Biomedical Physics, Department of Physics, Munich Institute of Biomedical Engineering, Technical University of Munich, 85748~Garching, Germany.}
\thanks{\hspace*{-1.8em}T.~Koehler is with Philips Research, 22335~Hamburg, Germany, and also with the Institute for Advanced Study, Technical University of Munich, 85748~Garching, Germany.}
\thanks{\hspace*{-1.8em}\mbox{H.-I.~Maack} was with Philips Medical Systems DMC GmbH, 22335~Hamburg, 
Germany.}
\thanks{\hspace*{-1.8em}J.~Herzen is with the Chair of Biomedical Physics, Department of Physics, Munich Institute of Biomedical Engineering, Technical University of Munich, 85748 Garching, Germany.}
\thanks{\hspace*{-1.8em}F.~Pfeiffer is with the Institute for Advanced Study, Technical University of Munich, 85748~Garching, Germany, with the Department of Diagnostic and Interventional Radiology, Klinikum rechts der Isar, Technical University of Munich, 81675~Munich, Germany, and also with the Chair of Biomedical Physics, Department of Physics, Munich Institute of Biomedical Engineering, Technical University of Munich, 85748~Garching, Germany.}
}
\maketitle

\begin{abstract}
X-ray dark-field imaging enables a spatially-resolved visualization of ultra-small-angle X-ray scattering.
Using phantom measurements, we demonstrate that a material's effective dark-field signal may be reduced by modification of the visibility spectrum by other dark-field-active objects in the beam.
This is the dark-field equivalent of conventional beam-hardening, and is distinct from related, known effects, where the dark-field signal is modified by attenuation or phase shifts.
We present a theoretical model for this group of effects and verify it by comparison to the measurements.
These findings have significant implications for the interpretation of dark-field signal strength in polychromatic measurements.
\end{abstract}
\section{Introduction}
\label{sec:introduction}

Contrast in conventional X-ray imaging is due to spatial variations in the fraction of X-ray flux attenuated by a sample.
However, the magnitude of achieved attenuation contrast is highly dependent on the samples' elemental composition:
Most biological tissues consist of light elements and thus only generate low attenuation contrast.

Besides attenuation, the refraction of X-rays induced by a sample can be imaged by X-ray phase-contrast imaging methods.
For light elements, variations in phase-shift interaction cross-sections are far higher than those of attenuation \cite{Momose2001}, enabling a multitude of applications for imaging of biological tissues.
Many X-ray phase-contrast imaging methods have high demands on spatial and/or temporal coherence, which limits their use to microfocus X-ray tubes or synchrotron sources \cite{Bravin2013}.

\subsection{Grating-based X-ray imaging}

Grating-based X-ray imaging \cite{David2002, Momose2003, Weitkamp2005}, has however been adapted to use with low-coherence X-ray sources \cite{Pfeiffer2006}.
In this imaging technique, an optical grating (modulation grating, $G_1$) is introduced in the beam which generates periodic intensity modulations (fringes) at certain downstream distances.
This is commonly achieved by exploiting the (fractional) Talbot effect\cite{Guigay2004}, but the shadow of an attenuating grating can also be used\cite{Huang2009,Olivo2007}.
Information about attenuation, phase-shift, and loss of fringe contrast (visibility) can be retrieved from the changes of the fringes caused by the sample.

In order to detect ultra-small-angle scatter (in the \SI{}{\micro\radian} range) with reasonable system lengths, fringe patterns smaller than typical detector pixel sizes are used.
This requires an attenuating grating (analyzer grating, $G_2$) in front of the detector, with a pitch matching that of the fringes.
The fringes' average distortion over each pixel can then be analyzed with the phase-stepping technique, i.e., by acquiring multiple images, while one of the gratings is laterally moved \cite{Weitkamp2005}.
This method is compatible with polychromatic X-ray sources\cite{Weitkamp2005}, and can be adapted to sources with low spatial coherence by introducing a grating near the X-ray source (source grating, $G_0$), which generates a periodic array of line sources\cite{Pfeiffer2006}.

Like attenuation and phase shift, visibility reduction can be interpreted as an imaging signal: besides the spectral effects mentioned below, the X-ray dark-field\cite{Pfeiffer2008} is a measure of ultra-small-angle X-ray scatter caused by a sample.
The signal's relationship to sample microstructure has been examined in a wide range of works \cite{Yashiro2010, Lynch2011, Strobl2014, Prade2015, Gkoumas2016, Ludwig2019, Graetz2020}.

\subsection{Spectral effects}
In analogy to the linear attenuation coefficient reconstructed by computed tomography (CT), an equivalent volumetric quantity can be defined for dark-field: the dark-field extinction coefficient \cite{Lynch2011}.
For a grating-based X-ray imaging setup using monochromatic X-rays, the interpretation of measured attenuation and X-ray dark-field signals is straightforward:
They are directly related to line integrals of the linear attenuation coefficient or dark-field extinction coefficient, respectively \cite{Bech2010}. As discussed later, the dark-field extinction coefficient also depends on the sample location within the setup \cite{vanStevendaal2013}.

Apart from Compton scattering, the interaction of X-rays with the sample does not affect their wavelength.
Unlike conventional X-ray imaging, the image formation process in a grating-based X-ray imaging setup is based on interference, and thus the presence of spatial correlations.
It can be shown that different temporal Fourier coefficients of a polychromatic wave field are always uncorrelated---consider e.g., the definition of cross-spectral density of a random process\cite{MandelWolf1995}.
Thus, no interference occurs between two fields with different photon energies.

It is therefore appropriate to calculate measured intensity separately for each photon energy, and finally integrate all intensities.
The case of polychromatic illumination can thus be interpreted as a superposition of the intensities at all photon energies.
Since the response of a grating-based imaging setup, as well as the sample properties, are dependent on photon energy, this leads to a complex, non-linear dependence of polychromatic transmittance and ultra-small-angle scatter on sample thickness.

For conventional X-ray imaging and CT, this effect is known as beam-hardening:
Since low-energy X-ray photons are usually attenuated more strongly than high-energy photons, the mean energy of photons transmitted through an object tends to be higher than that of the incident photons.
The attenuation of structures further downstream is then dominated by their properties at these higher energies, resulting in decreased attenuation. In CT, this may become apparent as a ``cupping'' artifact \cite{Hsieh2009}.

This effect is also present in polychromatic grating-based X-ray imaging in all three signal channels (attenuation, differential phase, and dark-field).
Additionally, while monochromatic dark-field and phase contrast are only a function of ultra-small-angle scatter and refraction, respectively, their polychromatic counterparts are affected by all three basic interactions: attenuation, ultra-small-angle scatter, and refraction.
In particular, the effect of beam-hardening on the dark-field signal is significant:
Measured visibility in a polychromatic setup is a \emph{weighted} average of photon-energy-dependent visibility (the visibility spectrum), and a change in spectral composition of the wave field affects the average's relative weights (harder X-rays are weighted more strongly downstream of an attenuating object).
This results in a change---often a decrease---of polychromatic visibility, which is not due to ultra-small-angle scatter.
Multiple approaches for quantifying and correcting this effect have been suggested, both for conventional grating-based imaging systems \cite{Yashiro2015, Pelzer2016} and dual-phase-grating systems \cite{Pandeshwar2020}.
\cite{Pelzer2016} has also examined the influence of wavelength-dependent refraction on the dark-field signal.
There have also been efforts to disentangle the impact of attenuation and edge-diffraction effects from dark-field images through algorithmic comparison of image content in the three grating-based imaging modalities~\cite{Kaeppler2014}.

For polychromatic signals however, we show here that the aforementioned weights in the visibility calculation also depend on the visibility spectrum.
This means that different dark-field extinction coefficients may be measured for the same material, depending on whether or not it is surrounded by other scattering materials---even if they induce little attenuation.

Similar to the linear attenuation coefficient, the magnitude of ultra-small-angle scatter is typically higher for lower photon energies.
Thus, the visibility at these energies is reduced more strongly by a sample, which shifts the center of the visibility spectrum towards ``harder'' X-rays.
We suggest to call this effect \emph{visibility-hardening}, due to its conceptual similarity to beam-hardening.

We present grating-based X-ray imaging measurements of a number of foam rubber and aluminum sheets. The foam rubber produces a strong dark-field signal but attenuates weakly, whereas aluminum attenuates strongly and produces a negligible dark-field signal.
The presence of visibility-hardening is apparent from the data (Section~\ref{sec:results}).

A theoretical model for the calculation of polychromatic visibility and dark-field is then introduced, and a regression of this model to the data is performed, confirming that visibility-hardening arises naturally from the model (Section~\ref{sec:theory}).

\section{Experimental procedures}
\label{sec:experiment}
\subsection{Imaging setup}
The setup used for the experimental procedure has previously been introduced 
in\cite{Andrejewski2021}, and is a modified version of the setup described 
in\cite{Gromann2017, Hellbach2017}.
It is a grating-based imaging setup designed for fringe-scanning acquisition\cite{Kottler2007} of large samples with a conventional X-ray source. The three-grating arrangement is mounted on a common frame which is pivoted about an axis through the X-ray tube focal spot. During fringe-scanning, the slot formed by the grating arrangement is moved across the entire detector area within $\SI{40}{\s}$.
For the measurements shown here however, the slot remains stationary, and a simple phase-stepping procedure is used.

The X-ray tube (MRC 200 0310 ROT-GS 1004, Philips Medical Systems DMC, Hamburg, Germany), high-voltage generator (Velara, Philips Medical Systems DMC, Hamburg, Germany), and flat-panel detector (Pixium RF~4343, Trixell, Moirans, France) are components used in medical imaging setups. 
The tube is operated in pulsed mode ($\SI{12}{\hertz}$, $\SI{20}{\ms}$ pulse duration) at $\SI{60}{kVp}$, using a tube current of \SI{600}{\mA}. The detector ($\SI{43 x 43}{\cm}$ FOV) is operated in a $3\times 3$ binning mode, yielding a pixel size of $\SI{444}{\um}$.
The $G_0$ and $G_2$ gratings were manufactured by the Institute of Microstructure Technology 
(IMT) at the Karlsruhe Institute of Technology (KIT), Karlsruhe, Germany, using the 
LIGA process \cite{Mohr2012}, while $G_1$ was made by 5microns GmbH, Berlin, 
Germany using deep-reactive ion-etching (DRIE).

\begin{figure}[htbp]
	\centering
	\includegraphics{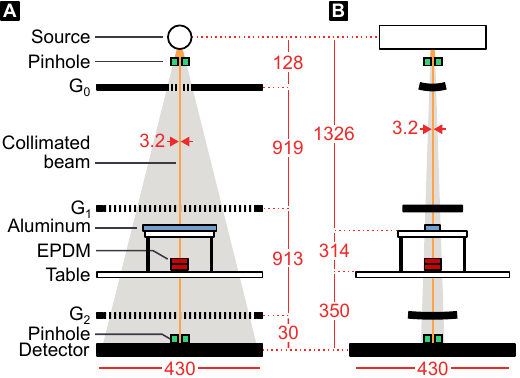}
	\caption{\small Schematic of the imaging setup with the placement of the imaged phantom. Not to scale, numbers are distances in mm. Gray: X-ray light cone during normal operation, orange: collimated X-ray light cone used for the experiment. A: Front view. B: Side view.}
	\label{fig:setup}
\end{figure}

$G_0$ and $G_2$ are absorption gratings, using gold as the absorbing material.
$G_0$ has a size of $\SI{30 x 70}{\mm}$, a nominal absorber height of approximately \SI{205}{\um}, and a duty cycle of $0.68$.
$G_1$ is a phase grating with a size of \SI{225 x 60}{\mm} and ridges with a height of \SI{59.1}{\um} etched into silicon, yielding a $\pi$ phase shift at \SI{40.2}{\keV}.
$G_2$ is an assembly consisting of eight grating tiles, with a nominal absorber height of about $\SI{200}{\um}$ and an area of $\SI{56 x 75}{\mm}$, yielding an active area of $\SI{448 x 75}{\mm}$.
Our calculations show, however, that the actual gold height for the two absorbing gratings in the examined area is considerably lower, close to \SI{150}{\um} (see section~\ref{subsec:verification}).
All gratings have the same pitch of $p=\SI{10}{\um}$, as they are used in a near-symmetric interferometer configuration ($G_0$--$G_1$~distance: $\SI{919}{\mm}$, $G_1$--$G_2$~distance: $\SI{913}{\mm}$).
$G_0$ and $G_2$ are bent to curvature radii comparable to their source distance (roughly \SI{250}{\mm} and \SI{2000}{\mm}, respectively \cite{Andrejewski2021b}), in order to minimize shadowing effects.
Setup geometry and dimensions are illustrated in Fig.~\ref{fig:setup}.

\subsection{Phantom construction and imaging}
\label{subsec:measurements}
An imaging phantom was constructed from uniform sheets of two materials:
aluminum was used for attenuation, whereas ethylene-propylene-diene monomer (EPDM) cellular foam rubber (DRG GmbH \& CoKG, Neumarkt am Wallersee, Austria) generated ultra-small-angle scatter.

Aluminum was chosen due to its homogeneity, ensuring that it generates no ultra-small-angle scatter.
While nanovoids in aluminum sheets have been found to generate a small-angle X-ray scatter signal \cite{Chaudhuri2013}, it is very weak, likely due to the extremely small size and volume fraction of these voids.
Unlike in a typical small-angle scattering experiment, where the scattering pattern is separated from the primary beam, scattered and unscattered radiation overlap spatially in X-ray dark-field imaging.
Thus, the dynamic range of a dark-field measurement is dictated by the total intensity, and not by the intensity of the scattered signal.
It is therefore very unlikely that aluminum with comparable properties to that in \cite{Chaudhuri2013} would be able to generate a measurable dark-field signal.

EPDM is a closed-cell foam rubber.
Its highly porous microstructure resembles alveolar structures in the lung \cite{Taphorn2020} and generates a strong dark-field signal.
Due to its low mass density and high air content, it attenuates X-rays only weakly.

Up to six \SI{10}{\mm} thick sheets of EPDM (\SI{80 x 80}{\mm}) were stacked on the setup's main sample table, located \SI{350}{\mm} above the detector.
The foam sheets were supplemented by a stack of up to $20$ layers of \SI{1}{\mm} thick aluminum sheets ($\SI{50 x 500}{\mm}$). They were placed on a smaller table, $\SI{314}{\mm}$ above the main sample table (see~Fig.~\ref{fig:setup}).

Although the impact of Compton scatter on the dark-field signal can be corrected~\cite{Kim2019,Urban2022}, we opted to minimize its influence through experimental means:
a pair of lead pinholes ($\SI{3.2}{\mm}$ diameter, $\SI{1}{\mm}$ thickness) were placed near the source and upstream of the $G_0$ grating. Their lateral positions were adjusted so that their openings fully overlapped.

Mean values and errors of logarithmic transmittance $-\ln \overline{T}$ and dark-field $-\ln \overline{D}$ were retrieved using conventional phase-stepping~\cite{Weitkamp2005}, see Appendix~\ref{sec:appendix_signal_retrieval} for details. As explained in more detail in Section~\ref{sec:theory}, we use overlined symbols to denote polychromatically measured quantities.

\section{Experimental results}
\label{sec:results}
\begin{figure}[htbp]
	\centering\includegraphics{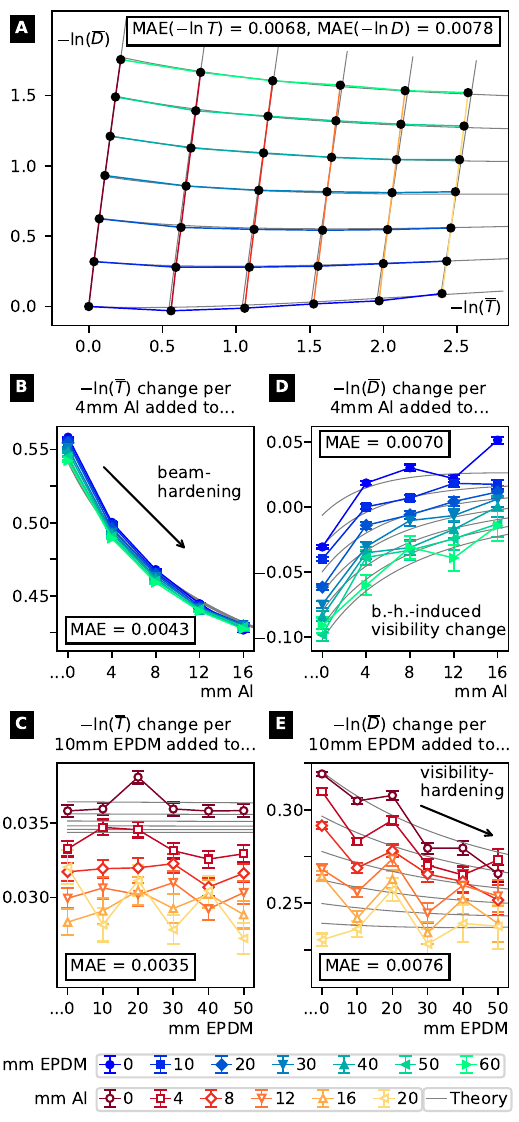}
	\caption{\small A:~Mean measured values of logarithmic dark-field and transmittance ($-\ln \overline{D}$, $-\ln \overline{T}$) for all thickness combinations of the phantom.
	Rows/columns of constant aluminum/EPDM thickness are color-coded in all subfigures (see legend). Fit of theoretical model (see Section~\ref{sec:theory}) to measurement data is shown in gray.
	B--E:~Signal level changes per added layer of aluminum/EPDM.
	B,~D:~Falling $-\ln \overline{T}$ per layer of aluminum characterizes the magnitude of beam-hardening.
	C:~Change in $-\ln \overline{D}$ per aluminum layer characterizes previously examined effect of beam-hardening on visibility~\cite{Yashiro2015,Pelzer2016}.
	E:~Decrease in $-\ln \overline{D}$ per EPDM layer illustrates the newly discovered visibility-hardening effect: the impact of preceding EPDM foam layers on the signal rivals that of aluminum, despite much weaker attenuation (\SI{60}{\mm}~EPDM $\approx$ \SI{1.6}{\mm}~Al). However, this effect is partially compensated by the positional dependence of autocorrelation length, as illustrated in Fig.~\ref{fig:sensitivity_variation}.
	Goodness of fit is reported for each subfigure as the mean absolute error (MAE) between measured and fitted signal levels.}
	\label{fig:point_grid}
\end{figure}

In Fig.~\ref{fig:point_grid}A, the mean values of $-\ln \overline{D}$ are plotted versus mean values of $-\ln \overline{T}$ for all aluminum\,/\,EPDM thickness combinations.
Furthermore, changes in $-\ln \overline{T}$ and $-\ln \overline{D}$ per added layer of aluminum or EPDM are shown in Fig.~\ref{fig:point_grid}B--E.

In a scenario with monochromatic radiation, constant measured autocorrelation length, and an absence of Compton scatter, the addition of an absorbing or scattering slab would lead to a constant increment of $-\ln \overline{T}$ or $-\ln \overline{D}$ respectively, regardless of the absolute signal levels \cite{Bech2010}.
All curves in Fig.~\ref{fig:point_grid}B--E would then be perfectly horizontal.
Since aluminum produces a negligible amount of ultra-small-angle scatter, curves in Fig.~\ref{fig:point_grid}D would be exactly zero.

In our measurements, we observe a number of deviations from this scenario:
Firstly, the added attenuation signal per layer of aluminum decreases with each layer (Fig.~\ref{fig:point_grid}B).
This well-known effect is due to stronger attenuation of low photon energies, shifting the  spectrum towards higher energies (beam-hardening), and thus a decreased attenuation of subsequent layers of material.
A similar effect is observable for the attenuation signal per EPDM layer (Fig.~\ref{fig:point_grid}C): it is decreased by preceding aluminum absorbers, but hardly by other EPDM layers due to their much weaker attenuation.

Fig.~\ref{fig:point_grid}D shows the change of $-\ln \overline{D}$ due to a layer of aluminum.
For many thickness combinations, this value is negative, meaning that the observed visibility with the additional aluminum is higher than without it.
This occurs since visibility depends on photon energy: as demonstrated in~\cite{Yashiro2015, Pelzer2016}, filtering the spectrum allows suppressing photon energies where visibility is low, leading to a higher average visibility.

Most importantly, Fig.~\ref{fig:point_grid}E shows the change in logarithmic dark-field signal per EPDM layer.
Two effects are observable: a decrease of added signal with the number of previous aluminum layers (different curves in Fig.~\ref{fig:point_grid}E), and a decrease with the number of previous EPDM layers (negative trend of these curves).
Note that both trends have a similar magnitude, but EPDM attenuates much more weakly than aluminum (\SI{60}{\mm}~EPDM $\approx$ \SI{1.6}{\mm}~Al), suggesting that beam-hardening is insufficient to explain the latter trend.
We attribute this decrease to the visibility-hardening effect.
In this measurement, visibility-hardening is partially compensated by the positional dependence of sampled autocorrelation length.
In section~\ref{sec:theory}, we discuss both effects in-depth, and show that the corresponding theory can explain them quantitatively.

\section{Theory}
\label{sec:theory}
In the following, we introduce a model to relate the polychromatically measured transmission and dark-field signals to the spectra of detected intensity and visibility.
To demonstrate the model's accuracy, we then apply it to the measured data.

\subsection{Calculation of polychromatic signals}
\label{subsec:theory}

\begin{figure}[h]
	\centering
	\includegraphics{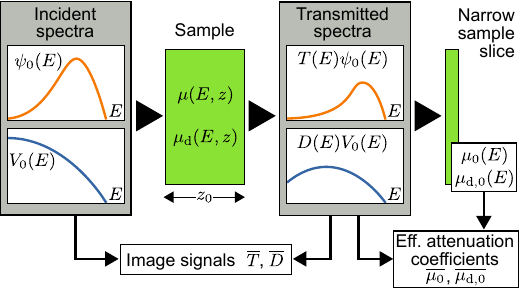}
	\caption{\small Outline of the calculations in Section~\ref{subsec:theory}: A sample with a linear attenuation coefficient $\mu(E,z)$ and dark-field extinction coefficient $\mu_\mathrm{d}(E,z)$ is imaged using polychromatic radiation with detector spectrum $\psi_0(E)$ and visibility spectrum $V_0(E)$.
		The object modifies the spectra according to the Beer-Lambert law, leading to downstream detector and visibility spectra $\psi(E)$, $V(E)$.
		The spectrally averaged transmittance and dark-field signals $\overline{T}$, $\overline{D}$ are calculated from the ratios of incident and transmitted flux and visibility spectra.
		Finally, the impact of the modified spectra on the effective attenuation coefficients of a downstream object with negligible thickness is examined.}
	\label{fig:theory_schematic}
\end{figure}

The approach for this calculation, as well as the used quantities, are outlined in Fig.~\ref{fig:theory_schematic}.
In conventional X-ray imaging, an object's transmittance is given by the fraction $T$ of incident radiation $S_0$ which passes through it.
This fraction is determined from two measurements with and without the object in the beam, and otherwise identical conditions.
For radiation with photon energy $E$, the amount of transmitted radiation $S$ is related to the object's linear attenuation coefficient $\mu$ via
\begin{equation}
\label{eq:lambert_beer}
T(E) = \frac{S(E)}{{S_0}(E)}  = \exp \left[-\int_0^{z_0} \mu(E,z) dz\right],
\end{equation}
with the beam propagating along the $z$ axis (object extends from ${z=0}$ to ${z_0}$).
The measured detector signals $S$ and $S_0$ in these measurements can be quantified by the radiation energy absorbed, or the registered number of X-ray photon absorption events.
$S$ and $S_0$ are proportional to the number of photons $n$, $n_0$ incident on the detector, if its response function $r = dS / dn$ does not depend on $n$.
$T$ is then independent of $r$, and thus of the detection mechanism.

This changes for polychromatic measurements: assuming independence of $r(E)$ from $n(E)$, the total signal is an integral over the signals from each photon energy interval $[E,E+dE]$:
\begin{equation}
\label{eq:signal_poly}
\overline{S_0} = \intl \psi_0(E) dE,\ 
\overline{S} = \intl \psi(E) dE,
\end{equation}
\begin{equation}
\label{eq:sigmas}
\psi_0(E) = \frac{\partial n_0(E)}{\partial E} r(E),\ \psi(E) = T(E) \psi_0(E),
\end{equation}
and $(\partial n_0 / \partial E) \, dE$ is the number of photons from the interval $[E, E+dE]$ incident on the detector in the reference image.


In an idealized model, the response function $r(E)$ of energy-integrating detectors is proportional to $E$ and the detector's quantum efficiency.
For real detectors, deviations occur due to the complex interactions of X-ray and visible photons with the scintillator material \cite{Heismann2012}.
Similarly, the response function of an idealized single-photon-counting detector depends only on its quantum efficiency \cite{Davidson2003}, but photon pileup, charge sharing, and X-ray fluorescence within the semiconductor material lead to a more complex behavior \cite{Koenig2012}.

Source and detector properties are included in the quantities $\psi_0(E)$ and $\psi(E)$. They are \emph{detector signals per photon energy interval} $[E, E+dE]$, for the reference and the sample measurement, respectively. They are used several times below.

As in \eqref{eq:signal_poly}, we use an overline to mark quantities as they would be measured in a polychromatic grating-based imaging setup with the spectra $\psi_0(E)$, $\psi(E)$.
Furthermore, we use the notation $\langle f \rangle_{g}$ to denote the mean of $f(E)$, weighted by $g(E)$, i.e.
\begin{equation}
\label{eq:wmean}
\langle f \rangle_{g} = \frac{\intl f(E) g(E) dE}{\intl g(E) dE},
\end{equation}
Transmittance in a polychromatic setup is then
\begin{align}
\label{eq:lambert_beer_poly}
\overline{T} &= \frac{\overline{S}}{\overline{S_0}}
\overset{\eqref{eq:signal_poly}, \eqref{eq:sigmas}}{=} \frac{\intl T(E) \psi_0(E) dE}{\intl \psi_0(E) dE}
\overset{\eqref{eq:wmean}}{=} \langle T \rangle_{\psi_0}.
\end{align}
Unlike in \eqref{eq:lambert_beer}, $r(E)$ does not cancel out, since it depends on $E$.
The effective linear attenuation coefficient $\overline{\mu_0}$ of an additional thin object with linear attenuation coefficient $\mu_0(E)=\mu(E,z_0)$ behind this object (see Fig.~\ref{fig:theory_schematic}) can be derived from \eqref{eq:lambert_beer_poly} via 
\begin{equation}
\label{eq:mut_eff}
\overline{\mu_0}
= - \frac{\partial \ln \overline{T}}{\partial \zt}
= \frac{1}{\overline{T}}\frac{\partial \overline{T}}{\partial \zt}
= \mean{\mu_0}{\psi},
\end{equation}
(see Appendix~\ref{sec:appendix_mueff}), i.e., it is the object's energy-dependent attenuation coefficient $\mu_0 (E)$, weighted by the incident detector spectrum $\psi_0(E)$, which has been filtered by the preceding absorber $T(E)$.
As lower photon energies are attenuated more strongly, $\psi(E)=\psi_0(E) T(E)$ shifts towards higher energies for increasing absorber thickness (beam-hardening).
The lower values of $\mu_0(E)$ at these energies lead to a decrease of $\overline{\mu_0}$ for increasing thickness of the preceding absorber.

X-ray dark-field imaging quantifies ultra-small-angle X-ray scatter by a sample via measuring the ratio of interferometric visibility with and without it: $D=V / V_0$.
The visibility characterizes the relative contrast of an interferometric fringe pattern. 
Given the highest and lowest detector signals $\smax$, $\smin$ in a given pattern, the visibility is $V = (\smax - \smin) / (\smax + \smin)$.
If the pattern's oscillations are symmetric around its mean intensity $S(E)$, it follows that
\begin{equation}
\smax(E) = S(E) [1+V(E)],\ \smin(E) = S(E)[1-V(E)].
\end{equation}
For monochromatic radiation, visibility also decreases exponentially as a function of sample thickness \cite{Strobl2014}:
\begin{equation}
\label{eq:df_spectral}
D(E) = \frac{V(E)}{V_0(E)} = \exp \left[-\int_0^{z_0} \mu_\mathrm{d}(E,z) dz\right].
\end{equation}
In analogy to the linear attenuation coefficient $\mu(E,z)$, the dark-field extinction coefficient $\mu_\mathrm{d}(E,z)$ characterizes this decrease \cite{Lynch2011}.
Under the assumption that no phase shift occurs, the polychromatically measured visibility $\overline{V}$ can be easily calculated from the spectral quantities:
In this case, all monochromatic detector signal modulations are in phase,
and when interpreting $\smax(E)$ and $\smin(E)$ as intensity per photon energy interval, their polychromatic equivalents are retrieved by integration:
\begin{align}
\overline{\smax} &= \intl \frac{\partial\smax}{\partial E} dE = \intl \!\!\! \psi(E) [1 + V(E)] dE,\\
\overline{\smin} &= \intl \frac{\partial \smin}{\partial E} dE = \intl \!\!\! \psi(E) [1 - V(E)] dE.
\end{align}
Given that, equivalently to the monochromatic case,\\
$\overline{V} = (\overline{\smax}-\overline{\smin})/(\overline{\smax}+\overline{\smin})$,
it follows that
\begin{equation}
\label{eq:lambert_beer_vis_poly}
\overline{V} = \frac{\intl V(E) \psi(E) dE}{\intl \psi(E) dE}
\overset{\eqref{eq:wmean}}{=} \langle V \rangle_{\psi}
= \langle V_0 D \rangle_{\psi_0 T}.
\end{equation}

The equivalent blank-scan visibility $\overline{V_0}$ is found by setting $T(E) = D(E) = 1$. Thus: $\overline{V_0} = \langle V_0 \rangle_{\psi_0}$.
Finally, the polychromatic dark-field signal is
\begin{equation}
\label{eq:lambert_beer_df_poly}
\overline{D} = \frac{\overline{V}}{\overline{V_0}}
=\frac{\langle V \rangle_{\psi}}{\langle V_0 \rangle_{\psi_0}}
= \frac{\langle V_0 D \rangle_{\psi_0 T}}{\langle V_0 \rangle_{\psi_0}}
\overset{(*)}{=} \frac{\mean{DT}{V_0 \psi_0}}{\left< T \right>_{\psi_0}}
\overset{\eqref{eq:lambert_beer_poly}}{=} \frac{\mean{DT}{V_0 \psi_0}}{\overline{T}},
\end{equation}
with $T$ and $D$ as given in \eqref{eq:lambert_beer} and \eqref{eq:df_spectral}.
(The related expression $\mean{V_0}{\psi} / \mean{V_0}{\psi_0}$ is used in \cite{Yashiro2015} to separately estimate the effect of beam-hardening).
The equality $(*)$ follows by applying \eqref{eq:wmean} and pairing the four integrals differently.
Comparison with \eqref{eq:lambert_beer_poly} demonstrates a direct relation of $\overline{D}$ with $\overline{T}$.

This can be understood intuitively through an alternative interpretation of \eqref{eq:lambert_beer_vis_poly}:
we can view the quantity $V \psi$ as the stepping curve's \emph{absolute} amplitude at energy $E$.
Thus, $V_0 \psi_0$ in \eqref{eq:lambert_beer_df_poly} could be seen as an ``amplitude spectrum''. Further, an object with transmittance $T$ and dark-field $D$ will reduce the stepping curve amplitude by $D T$. So the quantity $\mean{DT}{V_0 \psi_0}$ is simply the factor by which the sample reduces the stepping curve amplitude.

The effective dark-field extinction coefficient of a thin object at $z=z_0$ with attenuation coefficient $\mu_0(E)$ and dark-field extinction coefficient $\mu_\mathrm{d,0}(E)=\mu_\mathrm{d}(E,z_0)$ is (calculation see Appendix~\ref{sec:appendix_mudeff}):
\begin{align}
\label{eq:dfec_eff}
\overline{\mu_\mathrm{d,0}}
&= -\frac{\partial \ln \overline{D}}{\partial z_0}
\overset{\eqref{eq:lambert_beer_df_poly}}{=} \mean{\mu_\mathrm{d,0} + \mu_0}{V \psi} - \mean{\mu_0}{\psi} \\
&\overset{\eqref{eq:mut_eff}}{=}  \mean{\mu_\mathrm{d,0}}{V \psi} + \mean{\mu_0}{V \psi} - \overline{\mu_0}.
\end{align}
The dependence of $\overline{\mu_\mathrm{d,0}}$ on $\mu_0(E)$ illustrates the known effect of dark-field signals generated by attenuating objects.
The quantity ${\mean{\mu_0}{V \psi} - \overline{\mu_0}}$ characterizes its magnitude.
This term can be positive or negative, and becomes small e.g., if $\mu_0(E)$ or $V(E)=D(E) V_0(E)$ are roughly constant in the energy region where ${\psi(E)=T(E) \psi_0(E)}$ is nonzero.
This will be the case if $\psi_0(E)$  has a narrow bandwidth.

Additionally, the first term $\mean{\mu_\mathrm{d,0}}{V \psi}$ depends on $D(E)$, i.e., the ultra-small-angle-scatter of preceding materials.
As $D(E)$ is smaller for low energies, other scattering materials will shift the weighting function ${V(E) \psi(E)}$ towards higher energies, where $\mu_{\mathrm{d},0}(E)$ is usually lower, leading to a decreased $\overline{\mu_\mathrm{d,0}}$.

Going back to the previously mentioned alternative view using stepping curve amplitudes, we can rewrite \eqref{eq:dfec_eff} as
\begin{align}
	\overline{\mu_\mathrm{d,0}} &= \overline{\mu_{\mathrm{dt},0}} - \overline{\mu_0},\\
	\text{where } \overline{\mu_{\mathrm{dt},0}} &= \mean{\mu_\mathrm{d,0} + \mu_0}{V \psi} = -\frac{\partial \ln \mean{DT}{V_0 \psi_0}}{\partial z_0}.
\end{align}
Given our above interpretations for $DT$ and $V_0 \psi_0$, $\overline{\mu_{\mathrm{dt},0}}$ is simply the attenuation coefficient of the polychromatic stepping curve amplitude.
While this relationship is likely not of much practical use, it helps to explain the odd structure of~\eqref{eq:dfec_eff}.

The strength of the visibility-hardening effect can be quantified by $\partial \overline{\mu_\mathrm{d,0}} / \partial z_0$, i.e., the slope of the curves in Fig.~\ref{fig:point_grid}C and \ref{fig:point_grid}E. A lengthy calculation (see Appendix~\ref{sec:appendix_mud_deriv}) yields
\begin{align}
	\label{eq:vishardening_strength}
	&\frac{\partial \overline{\mu_\mathrm{d,0}}}{\partial z_0} = \operatorname{Var}_{\psi}\left( \mu_0 \right) - \operatorname{Var}_{V \psi} \left(\mu_\mathrm{d,0} + \mu_0 \right),\\
	\text{where } &\operatorname{Var}_w(X) = {\langle X^2 \rangle}_w - {\langle X \rangle}_w^2 = {\langle(X - \langle X \rangle _w)^2\rangle}_w,
\end{align}
with the weighted mean notation from~\eqref{eq:wmean}. This result implies that the magnitude of visibility-hardening is determined by the ``spread'' of the coefficient values [$\mu_\mathrm{d,0}(E) + \mu_0(E)$ and $\mu_0(E)$] across the energy range where the weighting functions [$V(E)\psi(E)$ and $\psi(E)$] are significantly greater than zero.
Furthermore, it shows that for primarily dark-field-active materials (${\mu_{d,0} \gg \mu_0}$), visibility-hardening will lead to a decrease of $\overline{\mu_\mathrm{d,0}}$, regardless of the spectra $\psi_0$ and $V_0$ (cf. Fig.~\ref{fig:point_grid}E).
In the absence of true dark-field, \eqref{eq:vishardening_strength} may be either positive or negative. Positive values (cf. Fig.~\ref{fig:point_grid}C) will be more common since $V\psi$ usually has a more narrow bandwidth than $\psi$.

\subsection{Verification with experimental data}
\label{subsec:verification}
In order to verify the findings from Section~\ref{subsec:theory},
we performed regression of \eqref{eq:lambert_beer_poly} and \eqref{eq:lambert_beer_df_poly} to the measurements shown in Fig.~\ref{fig:point_grid}.
We introduce an assumption about the energy-dependence of the foam's dark-field signal, and modify the general model to take the position-dependence of the setup's autocorrelation length $\xi$ into account.
We then determine the three remaining unknown model parameters (see Table~\ref{tab:values}) by least-squares regression of predicted signal levels ${-\ln \overline{T}}$, $-\ln \overline{D}$ to our measurements.
The signal levels predicted by the model are shown by the gray lines in Fig. 2.

Given a material's mass density $\rho$, and total interaction cross-section $\sigma^\mathrm{(tot)}(E)$ (expressed in $\mathrm{cm^2/g}$), its attenuation coefficient is
\begin{equation}
\mu(E) = \rho \, \sigma^\mathrm{(tot)}(E).
\end{equation}
The values for $\sigma^\mathrm{(tot)}$ were retrieved using the xraylib library~\cite{xraylib}.
EPDM is a polymer consisting of ethylene, propylene, and 0 to 12\% (weight) of a diene (Dicyclopentadiene, vinyl norbornene, ethylidene norbornene or 1,4-hexadiene)~\cite{Ravishankar2012}. This means EPDM is a mixture of carbon and hydrogen, with a carbon mass fraction of 85.6 to 86.2\% (we assumed 85.9\%).
Transmittance was then calculated with~\eqref{eq:lambert_beer}:
\begin{align}
	\label{eq:beer_lambert_T_sim}
	\nonumber
	T_{M,N}^{(\mathrm{calc})}(E) =
	\exp [ &- M \, \rho\sal \, h\sal \, \sigma \sal^\mathrm{(tot)}(E) \\
	 &- N \, \rho\sfr \, h\sfr \, \sigma\sfr^\mathrm{(tot)}(E) ],
\end{align}
where $\rho$ and $h$ are the densities and single-layer thicknesses of each material, ``FR'' here standing for ``foam rubber'', i.e., EPDM, and $N$ and $M$ are the number of layers of the respective materials.

On the other hand, energy-dependent dark-field extinction coefficients $\mu_\mathrm{D}(E)$ can not be directly derived from tabulated values:
As shown in \cite{Strobl2014}, the logarithmic dark-field is a function of both the macroscopic scattering cross-section $\Sigma$ and the real-space autocorrelation function $G(\xi)$ of the material's electron density:
\begin{equation}
\label{eq:df_autocorr}
-\ln D(E) = \int_{0}^{z_0} \Sigma(z,E) \left\{ 1 - G \left[z,\ \xi(z,E) \right] \right\} dz,
\end{equation}
where the autocorrelation length is given by $\xi(z,E) = hc (d_{2,0} + z)/(p_2 E)$, if the sample is downstream of $G_1$ (where $d_{2,0}$ is the distance between $z=0$ and $G_2$, $p_2$ is the $G_2$ pitch, $h$ is the Planck constant, and $c$ is the vacuum speed of light) \cite{Lynch2011}.
Thus, both $\Sigma$ and $G$ have an energy-dependence ($\Sigma$ is proportional to $E^{-2}$\cite{Yashiro2010}), but $G(\xi)$ is additionally dependent on the geometry of the sample's microstructure. It has been derived for various types of structures~\cite{Andersson2008}, but it is difficult to determine for a foam with mostly unknown structural parameters.

For the aluminum sheets, we assumed an absence of electron density inhomogeneities on the length scale of the sampled $\xi$ values, i.e., ${G \approx 1}$.
For the EPDM, we assumed that ${1 - G(\xi)}$ in \eqref{eq:df_autocorr} has a power-law dependence on photon energy, i.e., ${G \propto E^C}$, with an unknown exponent ${C<0}$.
This approach is a simplified variant of a model previously used to describe $G(\xi)$ for X-ray dark-field:
in~\cite{Yashiro2010}, a general model for random height fluctuations on surfaces, which was first introduced in \cite{Sinha1988}, is used which states that
\begin{equation}
	\label{eq:G_hurst}
	G(\xi) \approx \exp [ -(\xi / \xi_0)^{2H} ],
\end{equation}
where $\xi_0$ is the correlation length of phase fluctuations in the sample, and $H$ is the Hurst exponent (${0 < H < 1}$).
In that work, the authors find that $\xi_0$ and $H$ are closely related to the size and shape of the scattering structures, respectively.
Micro-CT measurements of EPDM foam (of the same batch as used in this work) have shown that its mean chord length, a measure for average cell size, and thus likely a good estimate for $\xi_0$, is around \SI{53}{\um}~\cite{Taphorn2020}.
The setup's sampled autocorrelation length, on the other hand, is around ${\xi = \SI{1.1}{\um}}$ (at the design energy of $E=\SI{40.2}{keV}$ and $d_2=\SI{350}{\mm}$), i.e., about two orders of magnitude smaller than $\xi_0$. We can thus use the approximation $\xi \ll \xi_0$ and apply it to~\eqref{eq:G_hurst} to yield
\begin{equation}
	G(\xi) \approx 1 - (\xi / \xi_0)^{2H},
\end{equation}
and therefore, inserting this in~\eqref{eq:df_autocorr},
\begin{align}
	\label{eq:df_autocorr_approx}
	\nonumber
	-\ln D(E) &\approx \int_0^{z_0} \Sigma(z,E) \left(\frac{\xi}{\xi_0}\right)^{2H} \!\! dz \\
	&\approx \int_0^{z_0} \Sigma(z,E) \left(\frac{hc d_2}{E p_2 \xi_0}\right)^{2H} \!\! dz
\end{align}
Since $\Sigma$ is proportional to $E^{-2}$, the full integrand then has a photon energy dependence with exponent $C=-2H-2$ ($-4 < C < -2$).
This qualitative behavior is also confirmed by energy-resolved dark-field measurements performed on different types of foam rubber (see Fig.~\ref{fig:spectra_and_epdm}B). These results show that $-\ln D(E)$ for these samples is proportional to $E^{-3}$ in a range of $\xi$ values comparable to those covered in the current experiment.
Thus, in the following we assume a value of ${C=-3}$ (corresponding to ${H=0.5}$).

However, since the height of the EPDM stack is not negligible compared to its distance from $G_1$, we must also take the positional dependence of the sampled $\xi$ value into account.
To do so, we assume that $\Sigma$ is independent of $z$ (i.e., uniform elemental composition and density), and introduce the information that
$d_2 = d_{2,0} + z$. (Fig.~\ref{fig:setup} shows that $d_{2,0}$, the distance from the $G_2$ grating to the bottom of the EPDM stack, is \SI{320}{\mm}). For an EPDM stack height of $l$, \eqref{eq:df_autocorr_approx} becomes
\begin{align}
	\nonumber
	-\ln D(E) &\approx \Sigma(E) \left(\frac{hc d_{2,0}}{E p_2 \xi_0}\right)^{2H}
	\underset{l_\mathrm{eff}(l)}{\underbrace{\int_0^l {\left(\frac{d_{2,0}+z}{d_{2,0}}\right)}^{2H} \!\! dz}},\\
	l_\mathrm{eff}(l) &= \frac{d_{2,0}}{2H+1}\left[ {\left( 1+\frac{l}{d_{2,0}}\right)}^{2H+1} - 1\right].
\end{align}
For ${l \ll d_{2,0}}$, the effective thickness ${l_\mathrm{eff}(l)}$ is  approximately $l$, but for larger values, it exceeds $l$.
The dark-field signal of $N$ EPDM sheets is thus
\begin{align}
\label{eq:beer_lambert_D_sim}
\nonumber
D_{N}^{(\mathrm{calc})}(E) &= \exp \left[ -\Sigma(E) \left(\frac{hc d_{2,0}}{E p_2 \xi_0}\right)^{2H} l_\mathrm{eff}(N l_0) \right]\\
&= \exp \left[ B\sfr \left(\frac{E}{E_0}\right)^{-2H-2}  \frac{l_\mathrm{eff}(N l_0)}{l_0} \right],
\end{align}
with $H=0.5$, $E_0 = \SI{40}{\keV}$, and ${l_0 = \SI{10}{\mm}}$ the thickness of a single EPDM sheet.
In order to group all non-energy-dependent terms, we have here introduced the (unitless) quantity
\begin{equation}
	B_\mathrm{FR} = \Sigma {\left(\frac{E}{E_0}\right)}^{2} {\left( \frac{hc d_{2,0}}{E_0 p_2 \xi_0} \right)}^{2H} l_0.
\end{equation}
Note that $B_\mathrm{FR}$ is independent of $E$ since $\Sigma$ is proportional to $E^{-2}$.
Thus, $B\sfr$ encodes the material's overall scattering strength (the dark-field per foam sheet at a photon energy of $E_0$), and ${l_\mathrm{eff}(N l_0) / l_0}$ is an effective number of EPDM sheets, corrected for the spatial dependence of $\xi$ (thus, slightly higher than $N$).

The X-ray tube spectrum was determined with simulations in PENELOPE / PENEPMA \cite{penelope} for a tungsten target and a take-off angle of \SI{11}{\deg}, the X-ray tube's anode angle.
The calculated spectrum is in good agreement with other tabulated tungsten anode spectra, such as the IPEM report~78 \cite{IPEM78}, and the SpekPy package \cite{Poludniowski2021}.

Exact values for tube filtration were not available, but from a schematic in \cite{Behling1990}, we estimated an approximate filtration of \SI{1}{\mm} of beryllium and \SI{9}{\mm} of transformer oil (assumed empirical formula \ce{CH2}, ${\rho=\SI{0.8}{\g \per \cm\cubed}}$). The tube's aluminum filtration was determined by regression of the theoretical signal formation model to the measurement data (see Section~\ref{sec:theory}), yielding a value of \SI{5.0}{\mm}.
Filtration due to the gratings was applied in addition to the preceding filters, considering their duty cycles and substrate thicknesses.
Energy-dependent detection efficiency was modeled using the degree of X-ray attenuation of a \SI{600}{\um} layer of CsI (the detector's scintillator material).
We also included filtration due to the two sample tables, i.e., $\SI{10}{\mm}+\SI{5}{\mm}$ of acrylic (polymethyl methacrylate).
The shape of the effective spectrum $\psi_0(E)$ is shown in Fig.~\ref{fig:spectra_and_epdm}A.

Visibility spectra $V_0(E)$ were simulated using a software package developed at the Chair of Biomedical Physics, Technical University of Munich. It calculates intensity modulations via Fresnel propagation of coherent wave fields \cite{Viermetz2022}. Additionally, the effects of a finite width of the source grating slots, and of the phase-stepping process are taken into account by appropriate convolution of intensity profiles. Visibility values are calculated from the intensity modulations using Fourier analysis.
Simulations were performed with photon energies from $5$ to $\SI{60}{\keV}$, in steps of $\SI{0.5}{\keV}$ (see Fig.~\ref{fig:spectra_and_epdm}A).
Initially, the height of the gold absorber of the source and analyzer gratings as provided by the manufacturer (\SI{200}{\um} and \SI{205}{\um}, respectively) were used for the simulation, but this resulted in deviations from experimental measurements. Further analysis showed that this could only be satisfactorily explained by gold heights being considerably lower in the region examined by the experiment. This is not too surprising, the gold height of LIGA-manufactured gratings with high aspect ratios is known to vary spatially.
The gold absorber height was thus added as a free parameter in our optimization (assuming identical heights for $G_0$ and $G_2$), yielding a value of \SI{147.4}{\um}.

Calculation of polychromatic transmittance and dark-field was done by solving \eqref{eq:lambert_beer_poly} and \eqref{eq:lambert_beer_df_poly}, and substituting $T$ and $D$ in these equations with $T_{M,N}^{(\mathrm{calc})}$ and $D_{M,N}^{(\mathrm{calc})}$ from \eqref{eq:beer_lambert_T_sim} and \eqref{eq:beer_lambert_D_sim}. This was done for all numbers of aluminum sheets ($M=0, 4, 8, \ldots, 20$) and EPDM sheets (${N=0, 1, \ldots, 6}$), applying the rectangle rule with a discretization of the integrands to \SI{0.5}{\keV} intervals.

The resulting simulated values $-\ln \overline{T}_{M,N}^\mathrm{(calc)}$, $-\ln \overline{D}_{M,N}^\mathrm{(calc)}$ for each thickness combination $(M,N)$ were then compared to the equivalent measured values $-\ln \overline{T}_{M,N}^\mathrm{(m)}$, $-\ln \overline{D}_{M,N}^\mathrm{(m)}$ (cf. Fig.~\ref{fig:point_grid}), and the sum of squared residuals
\begin{equation}
\label{eq:cost}
\mathcal{S} = \sum_{M,N}
\left[ \ln \overline{D}_{M,N}^\mathrm{(calc)} -
\ln \overline{D}_{M,N}^\mathrm{(m)} \right]^2 +
\left[ \ln \overline{T}_{M,N}^\mathrm{(calc)} -
\ln \overline{T}_{M,N}^\mathrm{(m)} \right]^2
\end{equation}
was minimized by variation of $B\sfr$, aluminum filter thickness $h_\mathrm{Al}$, and gold absorber height $h_\mathrm{Au}$.

As the fluctuations between adjacent data points in the ``differential'' signal plots (Fig.~\ref{fig:point_grid}B--E) significantly exceed what would be expected from the calculated statistical error levels (see error bars in the Figure), we infer that a significant, ``systematic'' error is present, which is probably mainly due to spatial variations in the composition (i.e., the number of interfaces and the effective thickness) of the EPDM layers.
This variation is probably present both between different EPDM layers, as well as different regions of the same layer. During the measurements, EPDM layers were added and removed, and although care was taken to place the layers in the same arrangement, this was probably not achieved with perfect precision.

Since these systematic errors usually appear to exceed the statistical errors (see error bars in Fig.~\ref{fig:point_grid}), we decided against weighting the residuals in \eqref{eq:cost} with the statistical uncertainties. Statistical errors vary strongly between data points (by factors of up to $10$), probably much more than the (unknown) systematic uncertainties. Weighting with statistical errors thus gives undue weight to the data points with low statistical errors (i.e., those at low phantom thicknesses), and allows improbably large deviations between fit and model for higher phantom thicknesses.
Conversely, estimating the magnitude of systematic deviations from the data is complicated and would require additional, uncertain assumptions.

Salient parameters for the regression procedure are listed in Table~\ref{tab:values}.
Regression results are overlaid onto the measured data as gray lines in all subfigures of Fig.~\ref{fig:point_grid}.

\begin{table}
	\caption{Quantities used for regression of the model to the data (gray lines in Fig.~\ref{fig:point_grid})}
	\label{tab:values}
	\setlength{\tabcolsep}{3pt}
	\begin{tabular}{|p{38pt}|p{130pt}|p{10pt}|p{15pt}|} 
		\hline
		Symbol& 
		\multicolumn{2}{p{142pt}|}{Quantity and origin $^{\mathrm{a}}$} &
		Value\\
		\hline
		$\rho\sal h\sal$&
		Area density of a \SI{1}{\mm} aluminum layer&
		M&
		\SI{0.270}{\g \per \cm \squared}\\
		$\rho\sfr h\sfr$&
		Area density of a \SI{10}{\mm} EPDM layer&
		M&
		\SI{0.164}{\g \per \cm \squared}\\
		$d_{2,0}$&
		Distance from $G_2$ to table surface&
		M&
		\SI{320}{\mm}\\
		$\sigma \sal^\mathrm{(tot)}(E)$&
		Total cross-section&
		T&
		---\\
		$\sigma \sfr^\mathrm{(tot)}(E)$&
		Total cross-section&
		T&
		---\\
		$h_\mathrm{Al}$ &
		Aluminum filter thickness of X-ray tube&
		R&
		\SI{5.03}{\mm}\\
		$B\sfr$&
		$-\ln D(E)$ of one \SI{10}{\mm} layer of EPDM at table height at $E_0=\SI{40}{\keV}$&
		R&
		$0.386$\\
		$h_\mathrm{Au}$ &
		Gold absorber height of the $G_0$ and $G_2$ gratings &
		R &
		\SI{147.4}{\um} \\
		\hline
		\multicolumn{4}{p{251pt}}{$^{\mathrm{a}}$ M~=~measured experimentally, T~=~retrieved from tabulated values, R~=~determined by regression of the model.}
	\end{tabular}
	\label{tab1}
\end{table}

\subsection{Simulation results}
Although the differences between theoretical results and measurement data exceed the determined measurement errors, the overall trends in the data (most obvious from Fig.~\ref{fig:point_grid}B--E) are well reproduced.
Among these four plots, the relative deviations for $-\ln \overline{T}$ increase per EPDM layer (Fig.~\ref{fig:point_grid}C) stand out. This is due to the very weak attenuation signal of the EPDM, compared to aluminum.
The strong relative deviations thus only contribute weakly towards the minimized quantity $\mathcal{S}$.

\begin{figure}
	\centering\includegraphics{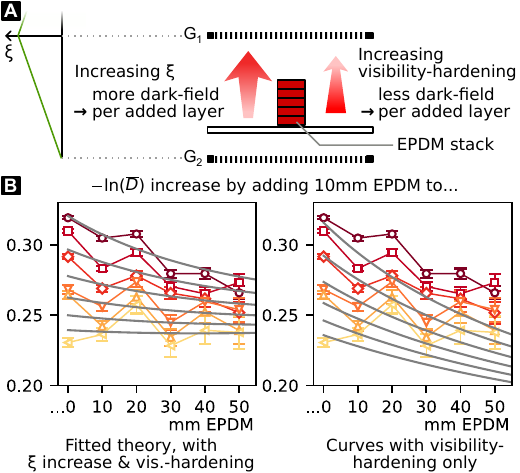}
	\caption{\small Competing effects of visibility-hardening and change of autocorrelation length $\xi$.
		A:~As $\xi$ is proportional to the sample's distance to the $G_2$ grating (far left), each successive EPDM foam layer is sampled at a larger $\xi$ value, \emph{increasing} the amount of dark-field per layer. Simultaneously, visibility-hardening by the other EPDM layers leads to a \emph{decrease} of dark-field per layer.
		B:~Impact of $\xi$ increase.
		Left: fit to data (like Fig.~\ref{fig:point_grid}E) including \emph{both} effects.
		Right: equivalent curves \emph{without} the $\xi$ increase, demonstrating the high impact of visibility-hardening. Note that the theory curves on the right are generated with ``frozen'' fit parameters $h_\mathrm{Al}$, $B_\mathrm{FR}$, $h_\mathrm{Au}$, i.e., these were not found by regression.}
	\label{fig:sensitivity_variation}
\end{figure}

The visibility-hardening effect is clearly observable at least for low aluminum thicknesses in Fig.~\ref{fig:point_grid}E. We find that the effect is somewhat neutralized by the increase in sampled autocorrelation length $\xi$ for every added EPDM layer. This effect is illustrated in Fig.~\ref{fig:sensitivity_variation}: if the variation of $\xi$ is neglected for the theoretical calculations [replacing $l_\mathrm{eff}(N l_0) / l_0$ by $N$ in \eqref{eq:beer_lambert_D_sim}], the amount of dark-field signal per EPDM layer continues to decrease with the overall amount of foam.

\begin{figure}
	\centering\includegraphics{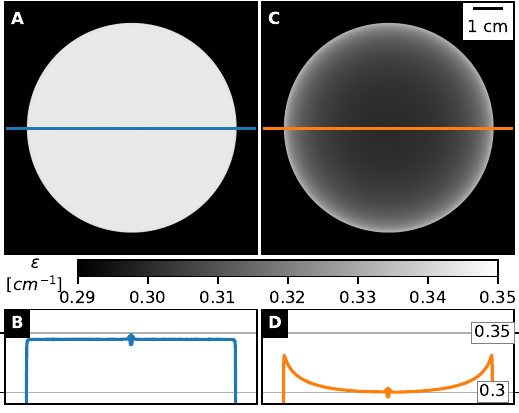}
	\caption{\small
		Effect of beam-hardening and visibility-hardening on tomographic dark-field data. Dark-field images of an EPDM cylinder with a diameter of $70\,\mathrm{mm}$ were simulated, and tomograms were retrieved by filtered backprojection.
		A:~Dark-field tomogram simulated with beam-hardening, but without visibility-hardening, and neglecting position-dependence of measured autocorrelation length $\xi$.
		B:~Line plot through A, exhibiting negligible cupping.
		C:~Dark-field tomogram with the assumption of $D(E)\propto E^{-3}$ (see Fig.~\ref{fig:spectra_and_epdm}B), and position-dependent $\xi$ as in the presented setup.
		D:~Line plot through C, exhibiting strong cupping.
	}
	\label{fig:ct}
\end{figure}

This effect can be even more clearly illustrated by simulating a tomographic measurement.
As a sample we assume a cylinder with a diameter of $7\,\mathrm{cm}$, composed of the same type of EPDM foam used in our measurements. Using \eqref{eq:lambert_beer_df_poly}, we generate a dark-field projection image of the cylinder, expand it into a sinogram, and apply filtered backprojection to generate a dark-field tomogram. For comparison, we then modify our model to exclude visibility-hardening (by assuming that dark-field signal $D(E)$ of EPDM is independent of photon energy $E$), and also exclude the position-dependence of measured autocorrelation length $\xi$. The two results are illustrated in Fig.~\ref{fig:ct}. While beam-hardening alone generates a perfectly flat reconstruction, a considerable cupping artifact, reminiscent of beam-hardening in conventional CT, appears due to the combination of visibility-hardening and the position-dependence of $\xi$.

The deviations between fit and measurement are probably due to spatial fluctuations in the phantom's material properties (especially the EPDM). Between the measurements, the EPDM and aluminum layers were not placed in exactly the same lateral positions. Although the shown error bars also include fluctuations of signal levels between pixels in the region of interest, the small size of the region may have lead to an underestimation of fluctuations on a length scale beyond a few millimeters.

\section{Discussion}
In this work, we have shown measurements of attenuation and dark-field signal levels of a phantom with well-defined thicknesses of absorbing and ultra-small-angle-scattering material.
We find a nonlinearity in the dark-field signal levels, which we tentatively call \emph{visibility-hardening}, and which has (to our knowledge) not yet been described in literature.
We presented a theoretical model for spectrally-averaged transmittance and dark-field, and validate this model by comparison to measurement data.
The model includes both known effects (the influence of beam-hardening on attenuation and dark-field, and the spatial variation of electron density autocorrelation $\xi$), as well as the observed visibility-hardening effect.

Due to their geometry, the used phantom materials generated no differential phase-shift signal. Its impact on the dark-field signal was therefore not examined.
We have found the differential-phase modality to carry little useful information in lung imaging, but it is possible that the phase signal from bone may distort dark-field signal levels.

The presented work was first started on another dataset, acquired at an earlier iteration of the shown setup (see the first version of this preprint). It did not use the Talbot effect; instead, an attenuating $G_1$ grating was used, and a very short $G_1$--$G_2$ distance meant that the "shadow" of the $G_1$ grating was analyzed instead of a self-image. This lead to significantly higher visibility $V_0(E)$ for a wide range of photon energies than a comparable Talbot-Lau setup. Furthermore, polyoxymethylene was used as an attenuating material instead of aluminum, and image data was acquired in a fringe scan without pinholes.

This older dataset appears to show a greater magnitude of visibility-hardening, which may be due to the different shape of the visibility spectrum.
However, it was found that the measurements were affected by unexpected effects (Compton scatter and/or dark-field caused by the attenuating material), whose magnitude was not controlled by the measurement procedure, making a clear distinction and quantification of the visibility-hardening effect difficult.
The procedure was thus repeated while minimizing these secondary effects, leading to the results shown in this work.

In principle, the presented findings could be used to simultaneously correct beam-hardening and visibility-hardening:
Using measurements from a calibration phantom, measured dark-field signals could be ``re-linearized'', i.e., converting $-\ln \overline{D}$ to a quantity proportional to the thickness of a (macroscopically homogeneous) ultra-small-angle-scattering material.
This signal would more accurately express the ``scattering power'' of the sample, and produce greater contrast between regions of high and low dark-field signal (e.g. healthy and emphysematous regions of the lung).
We think that such a correction would be important for dark-field CT measurements, such as the recently presented translation of dark-field to a medical CT device \cite{Viermetz2022b}, possibly in addition, or an alternative to, related corrections already performed \cite{Viermetz2023}.
In the presence of strong visibility-hardening, dark-field ``cupping artifacts'', equivalent to those in conventional CT, could occur.

However, such a correction would require the use of phantom materials with an energy-dependent dark-field extinction coefficient $\mu_\mathrm{d}(E)$ comparable to that of the sample.
Finding these may be challenging and may require comparative spectral dark-field measurements.
On the other hand, it is possible that an approximate similarity (e.g., similar diameter of foam rubber cells and lung alveoli) allows a sufficiently precise correction.
An accurate treatment of the positional dependence of $\xi$ may however be challenging, as the $z$ position of dark-field-active structures is not known a priori.
For CT, an iterative procedure is conceivable which extracts this information from an intermediate reconstruction of the dark-field tomogram.

For setups with a focus on dark-field applications, we thus recommend estimating the magnitude of visibility-hardening, using e.g., \eqref{eq:lambert_beer_poly}, \eqref{eq:lambert_beer_df_poly} for absolute signal levels, \eqref{eq:mut_eff}, \eqref{eq:dfec_eff} for effective attenuation coefficients, or \eqref{eq:vishardening_strength} for a simple numerical quantity characterizing the importance of the effect. Ideally, we think the effect should be minimized through setup design, in order to minimize the importance of corrections.

Besides corrections, the presented model could also be used to relate polychromatically measured dark-field and attenuation signal levels to microstructural quantities.
For example, pig lung image data from \cite{Ludwig2019}, together with known information about energy-dependent flux, visibility, and attenuation could be used to extract information about the energy-dependent dark-field $D(E)$, and thus $G(\xi)$.

Spectral X-ray data could be beneficial for estimating and reducing the effect of ``hardening'' effects.
GBI with photon-counting detectors using multiple energy bins has been successfully performed \cite{Thuering2013}, and benefits for dark-field signal retrieval from such data have been demonstrated \cite{Pelzer2014, Sellerer2021}.
As photon-counting detectors are increasingly finding their way into X-ray imaging applications, the practical importance of our model may decrease.
However, characterization of these detectors remains challenging, and the model could contribute to determining energy response functions from polychromatic calibration measurements.

\appendix
\section{Signal retrieval for measurements in section \ref{sec:experiment}}
\label{sec:appendix_signal_retrieval}
The three signal channels were retrieved using conventional phase-stepping~\cite{Weitkamp2005}.
For both sample and reference images, the $G_0$ grating was translated in equidistant steps orthogonal to the grating lines, using a motorized stage.
From this data, information was retrieved using the model
\begin{align}
	\label{eq:phase_stepping}
	I_k &= S [1 + V \cos(\varphi - 2 \pi x_k / p)],\\
	\label{eq:phase_stepping2}
	I_{0,k} &= S_0 [1 + V_0 \cos(\varphi_0 - 2 \pi x_k / p)],\\
	x_k &= kpL/K \ \ (k=1, \ldots, K;\,K,L\in\mathbb{N};\,K>L),
\end{align}
where $I_k$ and $I_{0,k}$ are the intensities in the $k$-th step in a given image pixel with or without the sample, respectively. Meanwhile, $p$ and $x_k$ represent pitch and $k$-th lateral position of the stepped grating ($G_0$).
The stepping curves' mean intensity, visibility, and moir\'e fringe phase ($S$, $V$ and $\varphi$ for the sample, $S_0$, $V_0$, and $\varphi_0$ for the reference) are found by least-squares regression of the model in \eqref{eq:phase_stepping}, \eqref{eq:phase_stepping2} to the measurements.

Reference and sample phase-stepping measurements were performed over $L=5$ grating periods in $K=11$ steps. A total of $20$ frames ($\SI{60}{\kV}$, $\SI{600}{\mA}$, $\SI{20}{\ms}$ exposure) were collected per step in order to achieve sufficient statistics.

Within \SI{91}{\min}, a phase stepping was recorded for each combination of EPDM thickness ($0,10,\ldots,\SI{60}{\mm}$) and aluminum thickness ($0, 4, \ldots, \SI{20}{\mm}$). Three reference measurements were acquired ($0$, $77$, and \SI{103}{\min} after the start). Reference values $S_0$, $V_0$, and $\varphi_0$ were calculated for each sample measurement by linear temporal interpolation over the reference measurements.
Transmittance $\overline{T}$, dark-field $\overline{D}$, and differential phase shift $\overline{\Delta \varphi}$ were calculated as
\begin{equation}
	\overline{T} = S / S_0, \ \overline{D} = V / V_0, \ \overline{\Delta \varphi} = \varphi - \varphi_0.
\end{equation}
Only pixels near the pinhole center (with ${>90\%}$ of peak intensity) were included to avoid penumbral effects, yielding a region of interest of $23$ pixels ($4.53\,\mathrm{mm}^2$ in the detector plane).
Mean values and errors of logarithmic transmittance and dark-field ($-\ln \overline{T}$, $-\ln \overline{D}$) were calculated across the region of interest. 

\section{Effective linear attenuation\\coefficient $\overline{\mu_0}$}
\label{sec:appendix_mueff}
\noindent Here we show that the effective linear attenuation coefficient $\overline{\mu_0}$, defined by the change in logarithmic transmittance with sample thickness $z_0$, is given by a weighted mean of the energy-dependent attenuation coefficient $\mu_0(E)$.
Using \eqref{eq:lambert_beer} and \eqref{eq:sigmas}, we find that:
\begin{align}
	\label{eq:dsigma_dz0}
	\frac{\partial \psi}{\partial z_0}
	\overset{\eqref{eq:sigmas}}{=} \frac{\partial T}{\partial z_0} \psi_0
	\overset{\eqref{eq:lambert_beer}}{=} - \mu(E,z_0) T \psi_0 \overset{\eqref{eq:sigmas}}{=} -\mu_0 \psi,
\end{align}
where $\mu_0 = \mu(E,z_0)$.
With this, we calculate the derivative of the negative logarithm of \eqref{eq:lambert_beer_poly} w.r.t.~$\zt$:
\begin{align}
	\nonumber
	\overline{\mu_0}
	&= -\frac{\partial \ln \overline{T}}{\partial \zt}
	= -\frac{1}{\overline{T}} \frac{\partial \overline{T}}{\partial \zt}
	\overset{\eqref{eq:lambert_beer_poly}}{=}
	-\frac{\intl \psi_0 dE}{\intl \psi dE} \frac{\partial}{\partial \zt} \left[ \frac{\intl \psi dE}{\intl \psi_0 dE} \right] \\
	&= -\frac{\frac{\partial}{\partial \zt} \left[ \intl \psi dE \right]}{\intl \psi dE}
	\nonumber
	= -\frac{\intl \frac{\partial \psi}{\partial \zt} dE}{\intl \psi dE}
	\overset{\eqref{eq:dsigma_dz0}}{=} \frac{\intl \psi \mu_0 dE}{\intl \psi dE} \\
	\label{eq:mu_eff}
	&\overset{\eqref{eq:wmean}}{=} \mean{\mu_0}{\psi}.
\end{align}
\section{Effective dark-field extinction coefficient $\overline{\mu_{\mathrm{d},0}}$}
\label{sec:appendix_mudeff}
\noindent Here we perform the equivalent to the calculation in Appendix~I, for the effective dark-field coefficient $\overline{\mu_{\mathrm{d},0}}$.
The procedure is nearly equivalent, but more complex because both attenuation and ultra-small-angle scatter affect the polychromatic dark-field signal $\overline{D}(E)$.
In equivalence to \eqref{eq:dsigma_dz0}, we find that
\begin{align}
	\frac{\partial V}{\partial z_0}
	\overset{\eqref{eq:df_spectral}}{=} \frac{\partial D}{\partial z_0} V_0
	\overset{\eqref{eq:df_spectral}}{=} - \mu_\mathrm{d}(E,z_0) D V_0
	=-\mu_{\mathrm{d},0} V.
	\label{eq:dv_dz0}
\end{align}
Using \eqref{eq:dsigma_dz0} and \eqref{eq:dv_dz0}, we can calculate the effective dark-field extinction coefficient:
\begin{align}
	\nonumber
	\overline{\mu_{\mathrm{d},0}}
	&= -\frac{\partial \ln \overline{D}}{\partial \zt}
	= -\frac{1}{\overline{D}} \frac{\partial \overline{D}}{\partial \zt}
	\overset{\eqref{eq:lambert_beer_df_poly}}{=} \frac{-\overline{T}}{\mean{DT}{V_0 \psi_0}} \frac{\partial}{\partial \zt}\left[ \frac{\mean{DT}{V_0 \psi_0}}{\overline{T}} \right] \\
	\label{eq:mud_intermediate}
	&= \frac{\partial \overline{T} / \partial \zt}{\overline{T}}
	- \frac{\partial \mean{DT}{V_0 \psi_0} / \partial \zt}{\mean{DT}{V_0 \psi_0}}.
\end{align}
From \eqref{eq:mu_eff}, we know that
$\frac{\partial \overline{T} / \partial \zt}{\overline{T}}= - \mean{\mu_0}{\psi}$.
Calculation of\\${\partial \mean{DT}{V_0 \psi_0} / \partial \zt}$ requires the product rule:
\begin{align}
	\nonumber
	\frac{\partial \mean{DT}{V_0 \psi_0}}{\partial \zt}
	&= \frac{\partial}{\partial \zt} \left[ \frac{\intl D T V_0 \psi_0 dE}{\intl V_0 \psi_0 dE} \right] \\
	&=  \frac{\intl \frac{\partial}{\partial \zt} \left[D T\right] V_0 \psi_0 dE}{\intl V_0 \psi_0 dE}.
\end{align}
The derivative follows from \eqref{eq:dsigma_dz0}, \eqref{eq:dv_dz0}:
\begin{align}
	\frac{\partial}{\partial \zt} \left[ D T \right]
	&= \frac{\partial D}{\partial \zt} T + \frac{\partial T}{\partial \zt} D 
	= - \left[ \mu_{\mathrm{d},0} +  \mu_0 \right] D T.
\end{align}
Therefore, the second term in \eqref{eq:mud_intermediate} is
\begin{align}
	\nonumber
	&\frac{\partial \mean{DT}{V_0 \psi_0} / \partial \zt}{\mean{DT}{V_0 \psi_0}}\\
	\nonumber
	&= \frac{- \intl V_0 \psi_0 dE}{\intl D T V_0 \psi_0 dE}
	\frac{\intl \left[ \mu_{\mathrm{d},0} +  \mu_0 \right] D T V_0 \psi_0 dE}{\intl V_0 \psi_0 dE}\\
	\nonumber
	&= \frac{-\intl \! \left[ \mu_{\mathrm{d},0} +  \mu_0 \right] D T V_0 \psi_0 dE}{\intl D T V_0 \psi_0 dE}
	= -\mean{\mu_{\mathrm{d},0} + \mu_0}{D T V_0 \psi_0},
\end{align}
and thus,
\begin{equation}
	\overline{\mu_{\mathrm{d},0}}
	= \mean{\mu_{\mathrm{d},0} + \mu_0}{D T V_0 \psi_0} - \mean{\mu_0}{\psi}.
\end{equation}

\section{Visibility-hardening magnitude\\
	$\partial \overline{\mu_{\mathrm{d},0}} / \partial z_0$}
\label{sec:appendix_mud_deriv}
Due to length restrictions, we only give a sketch of the calculation. We start by differentiating \eqref{eq:dfec_eff} w.r.t. $z_0$, writing out the weighted averages of the form $\langle \cdot \rangle_x$ with integrals according to \eqref{eq:wmean}. Since~\eqref{eq:wmean} is a division of two integrals, we apply the quotient rule, which yields a lengthy equation involving several integrals over photon energy, and derivatives of such integrals w.r.t. $z_0$.
The order of differentiation and integration can be swapped, and since the integrands contain several quantities that depend on $z_0$, we apply the product rule.
We employ the identities
\begin{align}
	\psi(E,z_{0}) & =T(E,z_{0})\psi_{0}(E)\label{eq:psi}\\
	V(E,z_{0}) & =D(E,z_{0})V_0(E)\label{eq:V}\\
	\frac{\partial T(E,z_{0})}{\partial z_{0}} & =-\mu(E,z_{0})T(E,z_{0})\label{eq:deriv_T}\\
	\frac{\partial D(E,z_{0})}{\partial z_{0}} & =-\mu_{d}(E,z_{0})D(E,z_{0})\label{eq:deriv_D}
\end{align}
in order to rephrase everything in terms of the variables on the right hand sides of the above equations. This yields
\begin{align}
	\nonumber
	&\frac{\partial}{\partial z_{0}}\left[\frac{\int\left(\mu_{\mathrm{d,0}}+\mu_{0}\right)V\psi\,dE}{\int V\psi\,dE}\right] =\\
	&\int\left\{ \frac{\partial\left(\mu_{\mathrm{d},0}+\mu_{0}\right)}{\partial z_{0}}-\left(\mu_{\mathrm{d},0}+\mu_{0}\right)^{2}\right\} V\psi\,dE,\label{eq:int1}\\
	&\frac{\partial}{\partial z_{0}}\left[\int V\psi\,dE\right] =
	-\int\left(\mu_{\mathrm{d},0}+\mu_{0}\right)V\psi\,dE.\label{eq:int2}
\end{align}
\eqref{eq:int1}, \eqref{eq:int2} can be used to get an intermediate result, the derivative of the first term representing $\overline{\mu_{\mathrm{d},0}}$ in \eqref{eq:dfec_eff}:
\begin{align}
	\nonumber
	&\frac{\partial}{\partial z_{0}}
	\left[
		\mean{\mu_{\mathrm{d,0}}+\mu_{0}}{V\psi}
	\right]=\\
	&\left\langle \frac{\partial\left(\mu_{\mathrm{d},0}+\mu_{0}\right)}{\partial z_{0}}-\left(\mu_{\mathrm{d},0}+\mu_{0}\right)^{2}\right\rangle _{V\psi}\!\!\!\!\!+\ \left\langle \mu_{\mathrm{d},0}+\mu_{0}\right\rangle _{V\psi}^{2}.\label{eq:1st_term_solved}
\end{align}
A similar calculation can be performed for the derivative of the second term, $\mean{\mu_0}{\psi}$, yielding
\begin{align}
	\frac{\partial}{\partial z_{0}}
	\left[
		\mean{\mu_0}{\psi}
	\right] =
	\left\langle \frac{\partial\mu_{0}}{\partial z_{0}}-\mu_{0}^{2}\right\rangle _{\psi}+\left\langle \mu_{0}\right\rangle _{\psi}^{2}.\label{eq:2nd_term_solved}
\end{align}
As the final step, we combine \eqref{eq:1st_term_solved} and \eqref{eq:2nd_term_solved}, while also setting $\partial \mu_0 / \partial z_0 = \partial \mu_{\mathrm{d},0} / \partial z_0 = 0$ (i.e., assuming locally constant attenuation coefficients:
\begin{align}
	\nonumber
	\frac{\partial\overline{\mu_{\mathrm{d},0}}}{\partial z_{0}}&=
	\left\{\left\langle \mu_{0}^{2}\right\rangle _{\psi} - \left\langle \mu_{0}\right\rangle _{\psi}^{2}\right \} \\
	&-\left\{ \left\langle \left(\mu_{\mathrm{d},0}+\mu_{0}\right)^{2}\right\rangle _{V\psi}-\left\langle \mu_{\mathrm{d},0}+\mu_{0}\right\rangle _{V\psi}^{2}\right\}.
\end{align}
The two right-hand side terms have an obvious similarity to the expression $\mathrm{Var}(X)=\langle X^2 \rangle - {\langle X \rangle}^2$. We perform the equivalent of this expansion with weighted means, which gives the final result of \eqref{eq:vishardening_strength}:
\begin{align}
	\nonumber
	\mathrm{Var}_{W}(X) = \left\langle \left(X-{\left\langle X\right\rangle}_{W} \right)^{2}\right\rangle _{W}
	=\left\langle X^{2}\right\rangle _{W}
	-{\left\langle X\right\rangle}_{W}^{2}.
\end{align}
Note that with setting $\partial \mu_0 / \partial z_0 = 0$, \eqref{eq:2nd_term_solved} can be written as $-\mathrm{Var}_{\psi}(\mu_0)$, a convenient expression to estimate the magnitude of beam-hardening for a given spectrum and absorber.

\begin{figure}
	\centering\includegraphics{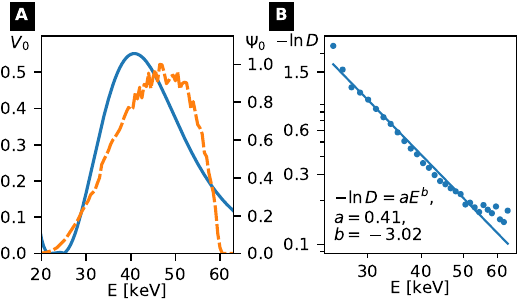}
	\caption{\small
		A: Simulated visibility spectrum $V_0(E)$ (solid) and detector spectrum $\psi_0$ (dashed) used for model regression.
		B: Energy-dependence of dark-field generated by EPDM.
		Measurement was performed with an X-123CdTe X-ray spectrometer (AMETEK Amptek, Bedford MA, USA). Regression shows good agreement with the power-law model from \eqref{eq:df_autocorr_approx} with an exponent near $-3$.}
	\label{fig:spectra_and_epdm}
\end{figure}

\section*{Acknowledgment}
The authors wish to thank Wolfgang Noichl for his help and fruitful discussions.
This work was carried out with the support of the Karlsruhe Nano Micro Facility (KNMF, www.kit.edu/knmf), a Helmholtz Research Infrastructure at Karlsruhe Institute of Technology (KIT). We acknowledge the support of the TUM Institute for Advanced Study, funded by the German Excellence Initiative.




\end{document}